\documentclass[epj,final]{svjour}
\usepackage{graphics,epsfig,rotating}
 \usepackage{amsmath}
 \usepackage{subfigure}
 \usepackage{placeins}
 \usepackage{amsmath}
 \usepackage{amssymb}
 \usepackage{wrapfig}
 \usepackage[]{tocbibind}
 \usepackage[all]{xy}

 \usepackage{ifthen}
 \usepackage{times}

\makeatletter
\newcommand\figcaption{\def\@captype{figure}\caption}
\newcommand\tabcaption{\def\@captype{table}\caption}\makeatother

\DeclareGraphicsExtensions{.jpg,.pdf,.png,.mps,.eps,.ps}  

\begin{document}

\DeclareGraphicsExtensions{.jpg,.pdf,.png,.mps,.eps,.ps}  

\title{Lambda hyperon production and polarization in collisions of p\,(3.5~GeV)\,+\,Nb}
\author{
\vskip-3bp
HADES collaboration
\\[2bp]
G.~Agakishiev$^{7}$, O.~Arnold$^{9}$, A.~Balanda$^{3}$, D.~Belver$^{18}$, A.\,V.~Belyaev$^{7}$, 
J.\,C.~Berger-Chen$^{9}$, A.~Blanco$^{2}$, M.~B\"{o}hmer$^{10}$, J.\,L.~Boyard$^{16}$, 
P.~Cabanelas$^{18,a}$, S.~Chernenko$^{7}$, A.~Dybczak$^{3}$, E.~Epple$^{9}$, L.~Fabbietti$^{9}$, 
O.\,V.~Fateev$^{7}$, P.~Finocchiaro$^{1}$, P.~Fonte$^{2,b}$, J.~Friese$^{10}$, 
I.~Fr\"{o}hlich$^{8}$, T.~Galatyuk$^{5,c}$, J.\,A.~Garz\'{o}n$^{18}$, R.~Gernh\"{a}user$^{10}$, 
K.~G\"{o}bel$^{8}$, M.~Golubeva$^{13}$, D.~Gonz\'{a}lez-D\'{\i}az$^{5}$, F.~Guber$^{13}$, 
M.~Gumberidze$^{5,16}$, T.~Heinz$^{4}$, T.~Hennino$^{16}$, R.~Holzmann$^{4}$, 
A.~Ierusalimov$^{7}$, I.~Iori$^{12,d}$, A.~Ivashkin$^{13}$, M.~Jurkovic$^{10}$, 
B.~K\"{a}mpfer$^{6,e}$, T.~Karavicheva$^{13}$, I.~Koenig$^{4}$, W.~Koenig$^{4}$, 
B.\,W.~Kolb$^{4}$, G.~Kornakov$^{18}$, R.~Kotte$^{6,*}$, A.~Kr\'{a}sa$^{17}$, F.~Krizek$^{17}$, 
R.~Kr\"{u}cken$^{10}$, H.~Kuc$^{3,16}$, W.~K\"{u}hn$^{11}$, A.~Kugler$^{17}$, A.~Kurepin$^{13}$, 
V.~Ladygin$^{7}$, R.~Lalik$^{9}$, S.~Lang$^{4}$, K.~Lapidus$^{9}$, A.~Lebedev$^{14}$, 
T.~Liu$^{16}$, L.~Lopes$^{2}$, M.~Lorenz$^{8,c}$, L.~Maier$^{10}$, A.~Mangiarotti$^{2}$, 
J.~Markert$^{8}$, V.~Metag$^{11}$, B.~Michalska$^{3}$, J.~Michel$^{8}$, C.~M\"{u}ntz$^{7}$, 
L.~Naumann$^{6}$, Y.\,C.~Pachmayer$^{8}$, M.~Palka$^{3}$, Y.~Parpottas$^{15,f}$, 
V.~Pechenov$^{4}$, O.~Pechenova$^{8}$, J.~Pietraszko$^{4}$, W.~Przygoda$^{3}$, 
B.~Ramstein$^{16}$, A.~Reshetin$^{13}$, A.~Rustamov$^{8}$, A.~Sadovsky$^{13}$, 
P.~Salabura$^{3}$, A.~Schmah$^{9,g}$, E.~Schwab$^{4}$, J.~Siebenson$^{9}$, Yu.\,G.~Sobolev$^{17}$,
S.~Spataro$^{11,h}$, B.~Spruck$^{11}$, H.~Str\"{o}bele$^{8}$, J.~Stroth$^{8,4}$, C.~Sturm$^{4}$, 
A.~Tarantola$^{8}$, K.~Teilab$^{8}$, P.~Tlusty$^{17}$, M.~Traxler$^{4}$, R.~Trebacz$^{3}$, 
H.~Tsertos$^{15}$, T.~Vasiliev$^{7}$, V.~Wagner$^{17}$, M.~Weber$^{10}$, C.~Wendisch$^{6,e,*}$, 
J.~W\"{u}stenfeld$^{6}$, S.~Yurevich$^{4}$, Y.\,V.~Zanevsky$^{7}$ 
}
\institute{
\mbox{ $^{1}$~Instituto Nazionale di Fisica Nucleare - Laboratori Nazionali del Sud,
95125~Catania, Italy}\\
\mbox{ $^{2}$~LIP-Laborat\'{o}rio de Instrumenta\c{c}\~{a}o e F\'{\i}sica
Experimental de Part\'{\i}culas , 3004-516~Coimbra, Portugal}\\
\mbox{ $^{3}$~Smoluchowski Institute of Physics, Jagiellonian University of Cracow,
30-059~Krak\'{o}w, Poland}\\
\mbox{ $^{4}$~GSI Helmholtzzentrum f\"{u}r Schwerionenforschung GmbH,
64291~Darmstadt, Germany}\\
\mbox{ $^{5}$~Technische Universit\"{a}t Darmstadt, 64289~Darmstadt, Germany}\\
\mbox{ $^{6}$~Institut f\"{u}r Strahlenphysik, Helmholtz-Zentrum Dresden-Rossendorf,
01314~Dresden, Germany}\\
\mbox{ $^{7}$~Joint Institute of Nuclear Research, 141980~Dubna, Russia}\\
\mbox{ $^{8}$~Institut f\"{u}r Kernphysik, Johann Wolfgang Goethe-Universit\"{a}t,
60438 ~Frankfurt, Germany}\\
\mbox{ $^{9}$~Excellence Cluster 'Origin and Structure of the Universe', 
85748~Garching, Germany}\\
\mbox{$^{10}$~Physik Department E12, Technische Universit\"{a}t M\"{u}nchen,
85748~Garching, Germany}\\
\mbox{$^{11}$~II.Physikalisches Institut, Justus Liebig Universit\"{a}t Giessen,
35392~Giessen, Germany}\\
\mbox{$^{12}$~Istituto Nazionale di Fisica Nucleare, Sezione di Milano,
20133~Milano, Italy}\\
\mbox{$^{13}$~Institute for Nuclear Research, Russian Academy of Science,
117312~Moscow, Russia}\\
\mbox{$^{14}$~Institute of Theoretical and Experimental Physics, 117218~Moscow, Russia}\\
\mbox{$^{15}$~Department of Physics, University of Cyprus, 1678~Nicosia, Cyprus}\\
\mbox{$^{16}$~Institut de Physique Nucl\'{e}aire (UMR 8608),
CNRS/IN2P3 - Universit\'{e} Paris Sud, F-91406~Orsay Cedex, France}\\
\mbox{$^{17}$~Nuclear Physics Institute, Academy of Sciences of Czech Republic, 
25068~Rez, Czech Republic}\\
\mbox{$^{18}$~LabCAF F. F\'{i}sica, Univ. de Santiago de Compostela, 
15706~Santiago de Compostela, Spain} 
\\
\mbox{ $^{a}$~also at Nuclear Physics Center of University of Lisbon, 1649-013 Lisboa, Portugal}\\
\mbox{ $^{b}$~also at ISEC Coimbra, 3030-199 Coimbra, Portugal}\\
\mbox{ $^{c}$~also at ExtreMe Matter Institute EMMI, 64291~Darmstadt, Germany}\\
\mbox{ $^{d}$~also at Dipartimento di Fisica, Universit\`{a} di Milano, 20133 Milano, Italy}\\
\mbox{ $^{e}$~also at Technische Universit\"{a}t Dresden, 01062~Dresden, Germany}\\
\mbox{ $^{f}$~also at Frederick University, 1036~Nikosia, Cyprus}\\
\mbox{ $^{g}$~now at Lawrence Berkeley National Laboratory, Berkeley, USA}\\
\mbox{ $^{h}$~now at Dipartimento di Fisica Generale and INFN, Universit\`{a} di Torino, 
10125 Torino, Italy}\\
\mbox{ $^{*}$~corresponding authors: r.kotte@hzdr.de, c.wendisch@hzdr.de}
}

\date{\vskip-6bp Received: \today}

\authorrunning{The HADES collaboration (G.~Agakishiev {\it et al.})}
\titlerunning{Lambda hyperon production and polarization in collisions of p\,(3.5~GeV)\,+\,Nb}

\abstract{\vskip-6bp
Results on $\Lambda$ hyperon production are reported for collisions of p\,(3.5~GeV)\,+\,Nb, 
studied with the High Acceptance Di-Electron Spectrometer (HADES) at SIS18 at GSI 
Helmholtzzentrum for Heavy-Ion Research, Darmstadt. 
The transverse mass distributions in rapidity bins are well described by Boltzmann 
shapes with a maximum inverse slope parameter of 
about 90\,MeV at a rapidity of $y=1.0$, i.e. slightly 
below the center-of-mass rapidity for nucleon-nucleon collisions, $y_{cm}=1.12$.  
The rapidity density decreases monotonically with increasing rapidity 
within a rapidity window ranging from 0.3 to 1.3. 
The $\Lambda$ phase-space distribution is compared with results of other 
experiments and with predictions of two transport approaches which are available 
publicly. None of the present versions of the employed models is able to fully reproduce the 
experimental distributions, i.e. in absolute yield and in shape. 
Presumably, this finding results from an insufficient 
modelling in the transport models of the elementary processes being relevant for 
$\Lambda$ production, rescattering and absorption. 
The present high-statistics data allow for a genuine two-dimensional investigation 
as a function of phase space of the 
self-analyzing $\Lambda$ polarization in the weak decay $\Lambda\rightarrow p \pi^-$. 
Finite negative values of the polarization in the order of $5-20\,\%$ are observed 
over the entire phase space studied.  
The absolute value of the polarization increases almost linearly with increasing transverse 
momentum for $p_t>300$\,MeV/c and increases with decreasing rapidity for $y < 0.8$. }

\PACS{{}25.75.Dw, 25.75.Gz}

\maketitle

\section{Introduction}\label{intro}
During the past two decades strangeness carrying particles which are produced 
in relativistic heavy-ion reactions at energies around the corresponding production thresholds  
in nucleon-nucleon (NN) collisions attracted strong attention  
from both experimental and theoretical sides 
\cite{Hartnack12,GiBUU,Fuchs04,CBMphysicsBook12,SengerStroebele99}
since their production and propagation 
are expected to provide insight into the nature of hot and dense nuclear matter. 
For instance, at energies available at SIS18/GSI Darmstadt densities up to three times that of 
ground-state nuclear matter and temperatures up to $\sim 100$\,MeV can be achieved. 
In the pioneering work of the KaoS collaboration at SIS18 on sub-threshold 
kaon production \cite{Sturm01} 
and based on comparisons to transport approaches \cite{Fuchs01,Hartnack06}
a soft equation of state (EoS) was extracted 
as predicted already long before \cite{AichelinKo85,LiKo95}. 
A strong focus was put on the study of in-medium properties of kaons and antikaons.   
Especially, their collective flow and spectral shapes at low transverse momenta are expected  
to be altered within the nuclear environment. The corresponding 
nuclear mean-field potential acting on the kaons and antikaons 
are predicted to be weakly positive ($\sim$ 20 to 40\,MeV) and strongly negative 
($\sim -50$ to $-80$\,MeV), respectively \cite{LiBrown_kpm98,CassBrat99}.  
Kaon and antikaon phase-space distributions  
\cite{Hartnack12,Kwisnia00,Foerster03,Merschmeyer07,Foerster07,Lotfi09,hades_K0_ArKCl} 
as well as azimuthal emission (flow) patterns of kaons 
\cite{Hartnack12,Crochet00,Uhlig05,Leifels10}     
support these predictions while the antikaon flow data are not conclusive yet. 
The tight interplay of experimental and and theoretical efforts   
revealed the importance of strangeness-exchange reactions 
for antikaon production ($YN\rightarrow {\bar K}NN$) and absorption 
(${\bar K}N\rightarrow Y\pi$). Also $\Lambda$ hyperons, being co-produced 
with kaons, are predicted to feel a weakly attractive potential ($\sim-30$\,MeV) 
\cite{LiKo96,LiBrown98,Wang99}. However, experimental data on the 
flow properties \cite{Ritman95} and on spectral shapes 
\cite{Merschmeyer07,Justice98,hades_Lambda_ArKCl} 
of hyperons produced in heavy-ion collisions near or below threshold are scarce. 
Evidently, not only nucleus-nucleus but also nucleon-nucleus collisions, 
though proceeding at densities not exceeding nuclear-matter ground-state density, 
may contribute to the discussion on medium modifications of meson and baryon  
properties. Here, valuable information has been collected by the KaoS 
collaboration, too \cite{Scheinast06}. The availability of such information 
is of great importance, since the reliable characterization 
of strangeness production and propagation in pA collisions is an indispensable 
prerequisite for the understanding of heavy-ion collisions. No experimental information is 
available on hyperon production in case of near-threshold nucleon-nucleus collisions which 
is assumed to serve as a link between elementary NN and heavy-ion collisions. 
Such data would allow for an improvement of the predictive power on the strangeness sector  
of state-of-the-art transport approaches which partially lack reliable input information 
on elementary processes. 

Parity conservation in strong interaction 
requires that the spin of produced $\Lambda$ hyperons is aligned 
perpendicularly to the production plane. 
In the parity-nonconserving weak process, $\Lambda \rightarrow p \pi^-$, the $\Lambda$ 
polarization causes a significant up-down asymmetry of the decay proton w.r.t. that plane. 
The $\Lambda$ polarization, surprisingly appearing in inclusive reactions 
with unpolarized beams and targets, was first observed in 1976 at Fermilab 
in collisions of p\,(300\,GeV)\,+\,Be \cite{Bunce76}. The observation of the negative 
polarization and the strong (almost linear) transverse-momentum dependence of its magnitude was 
rapidly confirmed in p\,(24~GeV/c)\,+\,Pt collisions at CERN-PS \cite{Heller77}, 
in p\,(400\,GeV)\,+\,Be 
again at the Fermilab neutral hyperon beam ($\Lambda$ and ${\bar \Lambda}$) \cite{Heller78}, 
in p\,(28.5\,GeV/c)\,+\,Ir at AGS \cite{Lomanno79}, and in p\,(12\,GeV)\,+\,W at KEK \cite{Abe83}. 
$\Lambda$ polarization measurements in pp interactions 
at the CERN intersecting storage rings \cite{Smith87} 
showed no obvious dependence on $\sqrt{s}$ which was varied from 31 to 62 GeV (which would 
correspond to 510 and 2050 GeV beam protons on a stationary proton target). 
However, the polarization 
was found to increase strongly with Feynman-$x$ ($x_F$) and with transverse momentum approaching 
-40\,\% when both quantities are maximum. Experiments with polarized beams followed soon, 
e.g. p\,+\,Be collisions at 13.3 and 18.5 GeV/c at the AGS \cite{Bonner88}. 
It is worth mentioning that all of these early experiments consisted of 
magnetic spectrometers set to fixed scattering angles. But more importantly, they 
suffered from a rather small solid-angle coverage    
resulting in a high degree of correlation between the 
transverse ($p_t$) and longitudinal ($x_F$ or rapidity) coordinates 
(e.g. in pN reactions at 450\,GeV at CERN-SPS \cite{Fanti99}). Any larger 
phase-space coverage implied a time-consuming setting of the corresponding spectrometer to 
different scattering angles with proper relative normalization as tried at Fermilab in 
p\,(400\,GeV) collisions on Be, Cu, and Pb targets with the main focus on the behaviour of the 
$\Lambda$ and ${\bar \Lambda}$ polarization at large transverse momentum \cite{Lundberg89}.
The first exclusive measurements of the $\Lambda$ polarization were performed with the E766 
experiment at the AGS in $pp\rightarrow K^+ \Lambda (\pi^+ \pi^-)^n$ (n=1-4) at 27.5\,GeV/c 
\cite{Felix96,Felix99}. $\Lambda$ polarization in $pp$ collisions is antisymmetric in 
$x_F$ by virtue of rotational invariance. Since the detector used had uniform acceptance in 
$x_F$, the same polarization magnitude was measured for positive and negative $x_F$. 
The behaviour of the polarization in $x_F$ and $p_t$ was found the same for all the studied 
reactions. 
First polarization studies of $\Lambda$ hyperons produced in heavy-ion collisions 
(10.7$A$\,GeV Au\,+\,Au) were reported by the E896 collaboration at BNL-AGS \cite{Bellwied02}. 
Their results revealed a dependence of the polarization on the transverse momentum and 
$x_F$ being consistent with previous measurements in pp and pA collisions.  
The polarization of $\Lambda$ hyperons produced inclusively by a $\Sigma^-$ beam of 
340\,GeV/c at CERN-SPS \cite{Adamovich04}, however, exhibited the striking feature of the 
mostly positive sign of the polarization which is opposite to what has been observed in 
$\Lambda$ production by protons or neutrons.  
The HERMES collaboration at DESY determined $\Lambda$ and ${\bar \Lambda}$ polarizations  
in quasireal photoproduction at the 27.6\,GeV positron beam of the HERA collider on an 
internal gas target \cite{Airapetian07}. Surprisingly, 
the $\Lambda$ polarization, averaged over the acceptance of the spectrometer, appeared positive,  
in contrast to almost all other experiments, while 
the ${\bar \Lambda}$ polarization appeared compatible with zero. 
A linear rise of the $\Lambda$ polarization magnitude 
with increasing transverse momentum was found similar to that observed in earlier experiments. 

At lower beam energies only few hyperon polarization data exist. The DISTO collaboration at 
SATURNE investigated strangeness production in elementary pp collisions with polarized proton beam 
at 3.67\,GeV/c \cite{Choi98}. First polarization transfer 
measurements for exclusive hyperon production reactions were reported in ref.\,\cite{Balestra99}. 
At COSY, the TOF collaboration studied $\Lambda$ 
polarization in the elementary $pp\rightarrow pK\Lambda$ reaction with polarized  
beams at 2.75 and 2.95\,GeV/c \cite{Pizzolotto07,Roeder11,Roeder12}. 
At higher energy and similar to DISTO \cite{Choi98} and E766 at AGS \cite{Felix96}, 
the experiments revealed a negative (positive) polarization in the beam (target) fragmentation 
region, while the polarization seemed to vanish at the lower energy. 
Besides the polarization of the $\Lambda$ hyperon, the $\Lambda$ analyzing power and the 
spin-transfer coefficient, also referred to as $\Lambda$ depolarization, could be determined. 
The lowest beam energy so far for which a $\Lambda$ polarization value could be determined 
amounts to 1.8\,GeV. For central nucleus-nucleus collisions of Ar+KCl at
this beam kinetic energy per incident nucleon, the streamer-chamber group at the BEVALAC reported an 
average $\Lambda$ polarization value of about $-0.10\pm0.05$ \cite{Harris81}.  
For recent reviews including parameterizations of various polarization dependences  
on kinematic quantities we refer the reader to refs.\,\cite{Felix99a,Siebert08,Nurushev2013}. 

While the $\Lambda$ polarization 
seems to be established as an experimental fact, its origin remains a mystery. 
Various models are proposed, assigned either to the quark-exchange 
\cite{Andersson79,DeGrand81,Zuo-tang97,Hui04,Yamamoto97,Kubo99} or to the meson-exchange picture 
\cite{Laget91,Soffer92,Sibirtsev98,Gasparian01,Sibirtsev06}.      
But, there is still no theoretical description which is able 
to explain the experimental observations consistently \cite{Felix99a}.  

Summarizing the experimental situation on $\Lambda$ production and polarization, we conclude 
that the high-acceptance spectrometer HADES \cite{hades_spectro} would be an appropriate  
experimental device allowing for I) the investigation of the $\Lambda$ phase-space 
distribution in proton-nucleus collisions and II) the study of a genuine two-dimensional 
(transverse, longitudinal) dependence of the $\Lambda$ polarization over a large 
phase-space region, feasible with one and the same apparatus setting.   

The present paper is organized as follows. In Sect.\,\ref{experiment} we give an overview 
of the HADES experiment on p\,+\,Nb collisions at 3.5\,GeV 
beam kinetic energy. 
We proceed with the data analysis in Sect.\,\ref{analysis}. 
In Sect.\,\ref{lambda_pid} we present the method 
to identify the $\Lambda$ hyperons from their weak decay into proton-$\pi^-$ pairs, 
while in Sect.\,\ref{lambda_phase_space} we describe 
the analysis chain for the extraction of experimental $\Lambda$ phase-space 
distributions and the corresponding predictions by transport models. In Sect.\,\ref{polarization} 
the results on the $\Lambda$ polarization and its phase-space dependence   
will be presented. Finally, in Sect.\,\ref{summary} we summarize our results. 

\section{The experiment}\label{experiment}
The experiment was performed 
with the {\bf H}igh {\bf A}cceptance {\bf D}i-{\bf E}lectron {\bf S}pectrometer (HADES)
at the Schwerionensynchrotron SIS18 at GSI, Darmstadt. HADES, although
primarily optimized to measure di-electrons \cite{HADES-PRL07}, offers excellent
hadron identification capabilities
\cite{hades_K0_ArKCl,hades_Lambda_ArKCl,PhD_Schmah,hades_kpm_phi,hades_Xi}
allowing for a profound correlation analysis.
A detailed description of the spectrometer is presented in ref.~\cite{hades_spectro}.
The present results are based on a dataset which was previously
analyzed with respect to e$^+$e$^-$ \cite{hades_epm_pNb} as well as to pion and $\eta$ 
production \cite{hades_pion_eta_pNb,Tlusty_Bormio_2012} in collisions of p\,+\,Nb at 3.5\,GeV;  
the production of $K^0$ mesons, 
focussing on the  $K^0$ phase-space distribution and its alteration due to 
the influence of a kaon-nucleon potential at nuclear-matter ground-state density,  
will be reported elsewhere. 
In the following we summarize the main features of the apparatus.

HADES consists of a 6-coil toroidal magnet centered on the beam axis and 
six identical detection sections located between the coils and covering 
polar angles from 18 to 85 degrees. The six sectors consist of hadron blind 
Ring-Imaging Cherenkov (RICH) detectors (not used for the present investigation), 
four planes of Multi-wire Drift Chambers (MDCs) for track reconstruction,   
and two time-of-flight walls, TOFino (polar angles $18^{\circ}<\theta<44^{\circ}$) and TOF 
($44^{\circ}<\theta<85^{\circ}$), supplemented at forward polar angles 
with Pre-Shower chambers. The TOF and TOFino+Pre-Shower detectors were 
combined into a Multiplicity and Electron Trigger Array (META). 
A reconstructed track in the spectrometer is composed of straight inner and outer track segments 
in the MDCs.  The pointing vector of the outer track segment is used for 
matching with a META hit. Possible trajectories through pairs of inner and outer 
track segments are combined to track candidates. 
A Runge-Kutta algorithm allows to calculate the momentum of each 
track candidate making use of the track deflection in the magnetic field 
between the inner and outer segments. The 
quality of the META-hit matching and the Runge-Kutta fitting 
(characterized by $\chi^2$ values) is used to 
create an ordered list of track candidates. The track candidate with the lowest 
product of both $\chi^2$ values is selected as the true 
track. Its segments and associated track candidates are then deleted from the candidate list. 
This procedure is repeated until no track candidates are left in the list. 

Particle identification of protons and $\pi^-$ mesons is based on the correlation of their 
momenta and energy loss in the MDCs. Two-dimensional cuts in the corresponding correlation plots  
are used to select the different particle species. For more details, 
e.g. the quality of kaon identification, see refs. \cite{PhD_Schmah,hades_kpm_phi}. 
Finally, the momentum calculated from the track curvature 
is corrected for the energy loss of the charged particles in the target, beam pipe and detector 
materials. 

In the present experiment, a proton beam of about $2\times 10^6$ particles per second with kinetic 
energy of 3.5~GeV  (corresponding to an excess energy w.r.t.   
the threshold for $\Lambda$ production in NN collisions of 
$\sqrt{s_{NN}}-\sqrt{s_{NN,\Lambda}}=0.63$\,GeV) 
was incident on a 12-fold segmented target of natural niobium ($^{93}$Nb). 
The choice of the target was the result of simulations aimed at optimizing the di-electron 
experiment \cite{hades_epm_pNb}, i.e. compromising on the ratio of the vector-meson production 
and the combinatorial background due to $\gamma$ conversion. 
However, the usage of this medium-size target is of advantage also for 
the present $\Lambda$ investigations, since the p-$\pi^-$ combinatorial background increases 
stronger with target mass than the $\Lambda$ yield which in turn is tightly connected to the 
production of the associated kaons \cite{Scheinast06}. 
The data readout was started by different trigger decisions 
\cite{hades_pion_eta_pNb}. For the present analysis, 
we employ only the data of the first-level trigger (LVL1, 
downscaled by a factor of three), 
requiring a charged-particle multiplicity $\ge 3$ in the TOF/TOFino detectors. 
We processed about $N_{ev}=3.2\times10^{9}$ of such LVL1 events. 
The total reaction cross section of $\sigma_{pNb}=(848\pm127)$\,mb  
is provided by measuring charged pions and by interpolating known
pion production cross sections \cite{hades_epm_pNb,Tlusty_Bormio_2012}. 

\section{Analysis} \label{analysis}
\subsection{$\mathbf{\Lambda}$ identification} \label{lambda_pid}
It is important to mention that $\Sigma^0$ hyperons decay almost exclusively 
into $\Lambda$'s via the decay $\Sigma^0 \rightarrow \Lambda \gamma$ 
(branching ratio $BR=100$\,\%, lifetime $c\tau=2.22\times10^{-11}$\,m \cite{PDG2012}), with  
the photon not being detected in the present experiment. 
Hence, throughout the paper, any ``$\Lambda$ yield''   
has to be understood as that of $\Lambda+\Sigma^0$. Correspondingly, in case of transport model  
simulations (cf. Sect.\,\ref{simulation}), where the individual particle species are known, 
the yields of $\Lambda$ and $\Sigma^0$ hyperons are summed up. 
Furthermore, we note that the actual $\Lambda$ polarization may be considerably larger than the 
measured one presented in Sect.\,\ref{polarization}, since $\Lambda$ hyperons from $\Sigma^0$ 
decay are expected to carry, on average, -1/3 the polarization of the $\Sigma^0$ \cite{Heller77}.  
The latter one is predicted \cite{Heller78,Andersson79} and measured \cite{Dukes87} 
to be opposite to that of the $\Lambda$. 

In the present analysis, we identify the $\Lambda$ hyperons through their weak decay 
$\Lambda \rightarrow p \pi^-$ ($BR=63.9$\,\%, $c\tau=7.89$\,cm \cite{PDG2012}), 
with the charged daughter particles detected in 
HADES \cite{hades_Lambda_ArKCl,hades_Xi}. 
The long lifetime of the $\Lambda$ causes a sizeable fraction   
of these particles to decay away from the primary vertex. The precision of the track 
reconstruction with HADES is sufficient to resolve these secondary vertices 
\cite{PhD_Schmah,hades_Xi}. For the selection of $\Lambda$'s, 
topological cuts are used. As a compromise between a high $\Lambda$ yield and a 
reasonable signal-to-background ratio ($> 0.1$ all over the investigated phase space), we choose 

i) a minimum value of the $\Lambda$ decay vertex distance to the primary vertex, $d_V>43$\,mm,  

ii) minimum values of the proton and $\pi^-$ shortest track distances to the primary vertex, 
$d_p>4$\,mm, $d_{\pi^-}>10$\,mm, and 

iii) an upper threshold of the proton-$\pi^-$ minimum track distance, $d_t<10$\,mm. 

\begin{figure}[!htb]
\begin{center}
\includegraphics[width=1.\linewidth,viewport=0 0 890 810]{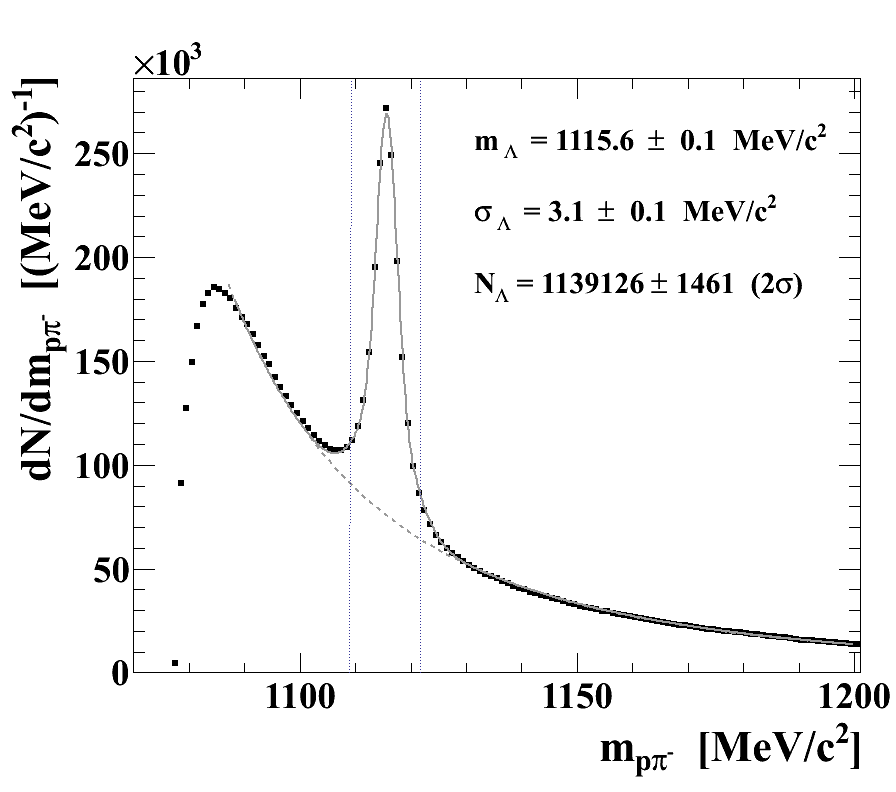}
\caption[]{Proton-$\pi^-$ invariant mass distribution (symbols) with the $\Lambda$-signal 
peak in the reaction p\,(3.5~GeV)\,+\,Nb. 
The full curve shows the result of a common fit with two Gaussians for the peak 
and a Tsallis function for the background (see text). The dashed curve represents the background. 
The vertical dotted lines bound the $2\sigma$ window for signal counting. 
\label{proton_piminus_inv_mass}}
\end{center}
\end{figure}

\noindent Here, the off-vertex cut i) is the main condition responsible for the extraction of 
a $\Lambda$ signal with good ($\sim$\,1) signal-to-background ratio. 
Figure\,\ref{proton_piminus_inv_mass} shows the invariant-mass distribution of all 
proton-$\pi^-$ pairs which pass the cuts listed above. To extract the $\Lambda$ yield, 
the invariant-mass spectrum is fitted, typically in the range from 1090 to 1200 MeV, 
with a combination of different functions describing both the signal and the combinatorial 
background of uncorrelated proton-$\pi^-$ pairs. 
The signal peak is parametrized by two Gaussians (with identical mean values but different 
widths to account for a certain broadening of the peak at its base), while the 
background is approximated by a Tsallis (q-exponential) function,  
$f_T(x) \propto (1-(1-q) x)^{1/(1-q)}$.  
Here, $x$ is a linear function of the invariant mass, $m_{p\pi^-}$, 
and $q$ is a shape parameter which may account for the phase-space limitation at large 
invariant masses, due to energy-momentum conservation.  
(Approximating instead the combinatorial background with the event-mixing technique, 
a quite similar quality of reproduction of the invariant-mass distribution 
over almost the entire mass ranges below and above the peak 
($\vert m_{p\pi^-}-m_{\Lambda}\vert>4\sigma_{\Lambda}$) is possible, 
with a slight exception of some underestimate of the data 
near ($m_{p\pi^-}<1085$\,MeV$/c^2$) to the kinematical limit, 
i.e. the sum of the proton and $\pi^-$ masses.) 
From the Gaussian fit to the peak, the pole mass is determined 
to be $1115.6\pm0.1$\,MeV$/c^2$, in very good agreement with the value of 
$1115.683\pm0.006$\,MeV$/c^2$ listed by the Particle Data Group \cite{PDG2012}. 
The peak width, which is a pure apparatus effect, is taken as the weighted average of  
the sigma widths of both Gaussians. It amounts to 3.1\,MeV$/c^2$, being only slighty larger 
than the corresponding widths measured in previous HADES analyses of the system 
Ar\,(1.76$A$\,GeV)\,+\,KCl \cite{hades_Lambda_ArKCl,hades_Xi} where, on average, lower momenta 
and hence tracks with higher curvature are involved. 
In total, for the cuts listed above and after background subtraction, 
about $1.1\times10^6$ $\Lambda$ hyperons were 
reconstructed within a $\pm 2\sigma$ window around the peak mass, 
with a mean signal-to-background ratio of 1.1. 

\subsection{$\mathbf{\Lambda}$ inclusive phase-space distribution} 
\label{lambda_phase_space}
\subsubsection{Experimental data} \label{lambda_phase_space_exp}
\begin{figure}[!htb]
\begin{center}
\includegraphics[width=1.\linewidth,viewport=0 0 780 930]{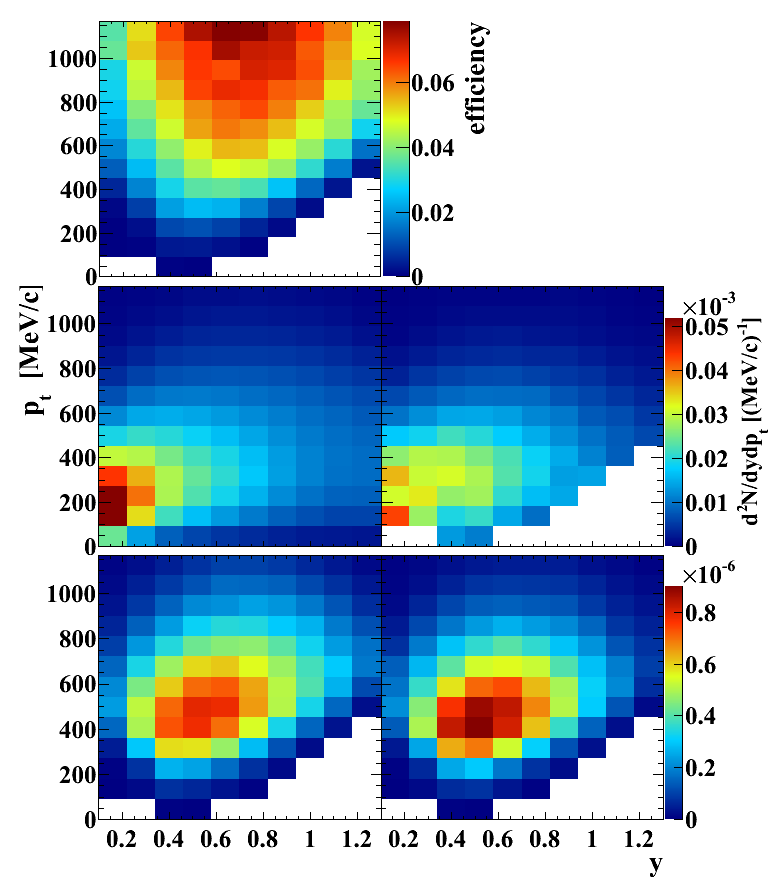}
\caption[]{Bottom right: Raw $\Lambda$ yield $d^2N/dp_t\,dy$. 
Bottom left: GEANT output of UrQMD simulations.  
Middle left: GEANT input (UrQMD output). 
Top left: Reconstruction efficiency matrix as derived from the ratio of the  
distributions below. Middle right: Experimental $\Lambda$ yield after correction 
with the efficiency matrix.  
\label{lambda_pty}}
\end{center}
\end{figure}

The bottom right panel of Fig.\,\ref{lambda_pty} shows the two-dimensional raw data yield 
(i.e. $\Lambda$ peak yields extracted from the proton-$\pi^-$ invariant-mass distributions 
after subtraction of the combinatorial background) as a function of transverse momentum 
and rapidity. Corrections for detector acceptance and reconstruction efficiency were performed 
with Monte-Carlo simulations involving, as appropriate event generator, the UrQMD 
transport approach \cite{UrQMD1,UrQMD2} and the GEANT \cite{GEANT} package accounting for the 
proper particle decays and the finite detector acceptance, granularity, resolution, etc. 
The minor missing experimental yield outside the $2\sigma$ window around the $\Lambda$-peak mass   
(cf. Fig.\,\ref{proton_piminus_inv_mass}) is well considered by applying the same cut 
to the simulation which exhibits a similar broadened peak base proved to be due to small-angle 
scattering in air and in the detector materials of the $\Lambda$ decay products. 
The bottom left panel of Fig.\,\ref{lambda_pty} 
displays the GEANT output, while the middle left panel shows the UrQMD output.  
The ratio of both distributions of simulated $\Lambda$ data delivers the corresponding 
reconstruction efficiency matrix (top left panel). Note that also the bias on 
the $\Lambda$ yield due to the effect of the LVL1 trigger is estimated with the 
help of UrQMD+GEANT simulations. An enhancement of 
the LVL1 triggered yield over the minimum-bias value of $1.53\pm 0.02$ has been found which 
is corrected for when determining the final (minium-bias) $\Lambda$ yield as a function of 
phase-space population. Within the given marginal uncertainty, 
the correction factor itself exhibits no dependence on phase space. 
Finally, after dividing -- for each phase-space cell -- the experimental 
raw data by the corresponding efficiency,  
the distribution corrected for acceptance, reconstruction efficiency, 
LVL1 trigger bias, and non-target interaction ($\simeq$17\,\% events w/o tracks), 
is derived and displayed in the middle right panel of 
Fig.\,\ref{lambda_pty}. Note that this phase-space distribution is the 
prerequisite for the two-dimensional investigation of the $\Lambda$ polarization 
presented in Sect.\,\ref{polarization}.  

\begin{figure}[!htb]
\begin{center}
\includegraphics[width=1.\linewidth,viewport=0 0 670 760]{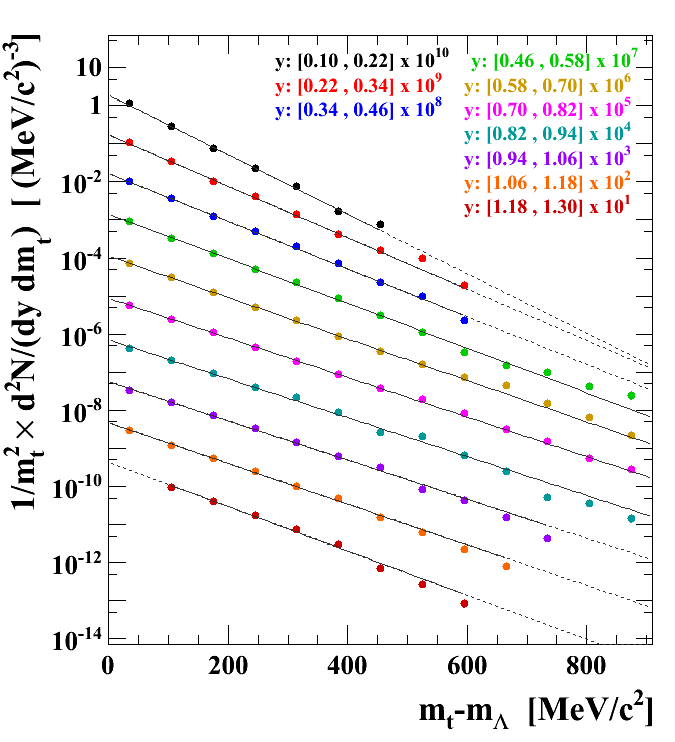}
\caption[]{Experimental $\Lambda$ yield distribution $m_t^{-2} d^2N/dm_t\,dy$ (symbols 
with error bars) in slices of rapidity as given in the legend. 
The lines represent fits with an exponential distribution according to Eq.\,(\ref{def_mt_distr}). 
Full lines cover the corresponding fit regions. Extended dashed lines 
display the part used for extrapolation.
\label{lambda_mt_fit}}
\end{center}
\end{figure}

To extrapolate into the unmeasured region at low transverse momentum, we follow the 
widely applicable and commonly used recipe of approximating the $p_t$ spectra by  
Maxwell-Boltzmann distributions. To do so, we start with the triple differential yield  
$d^3N/d^3p \propto \exp{(-E/T)}$. 
Here, $E=((p c)^2+(m_0 c^2)^2)^{1/2}=m_t c^2 \cosh y$ is the total energy,
$y=\tanh^{-1}(p_{\|}c/E)$ is the lab. rapidity, $m_t=((p_t/c)^2+m_0^2)^{1/2}$ is the transverse 
mass, $p_t$ ($p_{\|}$) is the transverse (longitudinal in beam direction) momentum, and $m_0$ 
is the rest mass of the particle of interest. Finally, $c$ gives 
the speed of light in vacuum. Transformation from spherical to cylindrical coordinates 
(being more appropriate for an almost rotational symmetric apparatus) and 
integration over the azimuthal angle delivers the two-dimensional yield 
\begin{equation}
\frac{d^2N}{m_t^2 dm_t dy} = C_B(y) \, \exp(-\frac{m_t c^2}{T_B(y)}). 
\label{def_mt_distr}
\end{equation}
Fits with a simple exponential to these transverse-mass spectra in slices of 
rapidity deliver constants, $C_B(y)$, and inverse (Boltzmann) slope parameters, $T_B(y)$. 
Figure\,\ref{lambda_mt_fit} shows the $\Lambda$ $m_t$ distributions for the 
indicated rapidity regions. The lines are Boltzmann fits to the data according to 
Eq.\,(\ref{def_mt_distr}).  For the determination of the $dN/dy$ distribution, 
the experimental data, where available, are integrated, and the yield in the unmeasured region 
is obtained from the fit.    

\begin{figure}[!htb]
\begin{center}
\includegraphics[width=1.\linewidth,viewport=0 0 630 640]{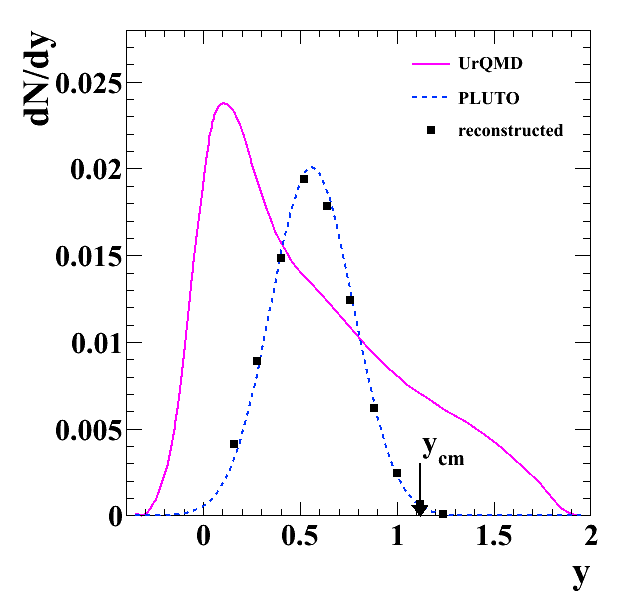}
\caption[]{Rapidity density distribution of Pluto simulated $\Lambda$ hyperons (arbitrarily 
scaled) before (dashed curve) and after (symbols) the full analysis 
chain as applied for experimental data. The full curve represents the corresponding spectral 
shape of the UrQMD simulations used for acceptance and reconstruction efficiency correction. 
The arrow indicates the center-of-mass rapidity in nucleon-nucleon collisions at 3.5\,GeV. 
\label{lambda_rap_self_consist}}
\end{center}
\end{figure}

Before we present the rapidity dependences of the inverse slope parameter, $T_B(y)$, and the 
rapidity density distribution, $dN/dy$, of $\Lambda$ hyperons produced in p+Nb collisions, 
we show the result of a self-consistency check, i.e. using UrQMD simulations for the acceptance 
and reconstruction efficiency corrections and analyzing another simulation similarly to    
the experimental data. For this purpose, we used the fireball option of the event generator 
Pluto \cite{Pluto}. Based on the expected experimental phase-space distribution, we 
generated $\Lambda$ hyperon events populating the phase space according to an isotropic 
thermal source with a temperature of $T=50$\,MeV centered at an average 
rapidity of $\langle y \rangle=0.56$. The Pluto events were tracked through GEANT to account for 
the detector response and the geometrical 
decay topology. The GEANT output was embedded into experimental events to account for a proper 
track environment and, finally, passed through the entire analysis chain with the 
correction matrix taken from UrQMD simulation.  
Figure\,\ref{lambda_rap_self_consist} shows the result for the rapidity distribution. 
Though the shapes of both simulations differ significantly, the initial shape of the 
phase-space distribution generated by Pluto is well recovered. Hence, we are convinced that  
our reconstruction method is based on solid grounds. 
Furthermore, we note that the experimental distributions given below represent the average 
of two analyses from two independent groups \cite{Wendisch14,Arnold13}. 
The given systematic errors comprise the slight differences of 
the corresponding analysis results and those from a variation of the cut values 
given in Sect.\,\ref{lambda_pid} within reasonable limits ($\pm 20\%$). The statistical errors 
are often smaller than the symbols displayed in the figures. 

\begin{figure}[!htb]
\begin{center}
\includegraphics[width=1.\linewidth,viewport=0 0 590 650]{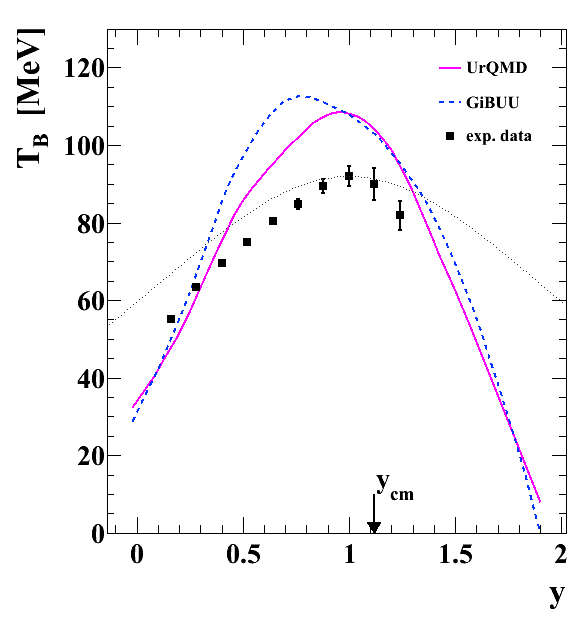}
\caption[]{Rapidity dependence of the experimental $\Lambda$ inverse (Boltzmann) slope 
parameter, $T_B(y)$, (symbols) as derived from the 
exponential fits in Fig.\,\protect\ref{lambda_mt_fit}. 
The error bars represent the systematic errors. The full and dashed curves display similar 
dependences resulting from corresponding fits to the spectra of the transport approaches 
UrQMD and GiBUU, respectively. 
The dotted curve represents the function $T_{B}(y)=92\,{\mathrm MeV}/\cosh(y-1.0)$. 
The arrow indicates the center-of-mass rapidity in nucleon-nucleon collisions at 3.5\,GeV. 
\label{lambda_TB}}
\end{center}
\end{figure}

Figure\,\ref{lambda_TB} presents the dependence of the $\Lambda$ inverse (Boltzmann) 
slopes on rapidity, 
$T_B(y)$ (with different transport model predictions overlaid, for discussion see 
Sect.\,\ref{simulation}). 
The error bars represent the uncertainties arising from the variations of the 
topological cuts to select the $\Lambda$ hyperons and of the borders of the  
Boltzmann fits applied to the $m_t$~($p_t$) distributions displayed in 
Fig.\,\ref{lambda_mt_fit}~(\ref{lambda_pt_spect_model}).
We find slope parameters, i.e. apparent transverse temperatures, 
from 55 to 92\,MeV with the maximum at a rapidity of $y_{max}=1.0$, 
that is below the center-of-mass rapidity of the nucleon-nucleon reference system, $y_{cm}=1.12$. 
Note that in symmetric heavy-ion collisions, e.g. of Ar\,+\,KCl at 1.76$A$\,GeV 
($y_{cm}=0.86$, HADES \cite{hades_Lambda_ArKCl}) and Ni\,+\,Ni at 1.93$A$\,GeV 
($y_{cm}=0.89$, FOPI \cite{Merschmeyer07}) at SIS18 or of Au+Au at 10.7$A$\,GeV at AGS  
($y_{cm}=1.60$, E896 \cite{Albergo02}), 
the rapidity dependence of the $\Lambda$ inverse slope parameter was found to 
follow well the thermal model prediction, i.e. a decline of the slope with increasing 
distance from mid-rapidity as $T_B=T/\cosh(y-y_{cm})$. 
Corresponding mid-rapidity values $T=(95.5\pm2)$, $(106\pm5)$, and $(237\pm5)$\,MeV have been 
reported in refs. \cite{hades_Lambda_ArKCl}, \cite{Merschmeyer07}, and \cite{Albergo02}, 
respectively. 
Note that, in a recent analysis of $\eta$ 
meson production in our reaction, p\,(3.5~GeV)\,+\,Nb, we found a similar symmetric rapidity 
dependence of the inverse slope, which could be parametrized with 
the above $1/cosh(y)$ dependence, however with the maximum located well below $y_{cm}$, 
i.e. $T_{B,\eta}(y)=84\,{\mathrm MeV}/\cosh(y-0.96)$ \cite{hades_pion_eta_pNb}. In contrast, 
the present rapidity dependence of the Boltzmann slope parameter for $\Lambda$ hyperons 
does not follow the isotropic thermal model prediction. 
Rather, it falls faster than $\propto 1/\cosh(y-y_{max})$, cf. dotted curve in 
Fig.\,\ref{lambda_TB}.  

\begin{figure}[!htb]
\begin{center}
\includegraphics[width=1.\linewidth,viewport=0 0 680 630]{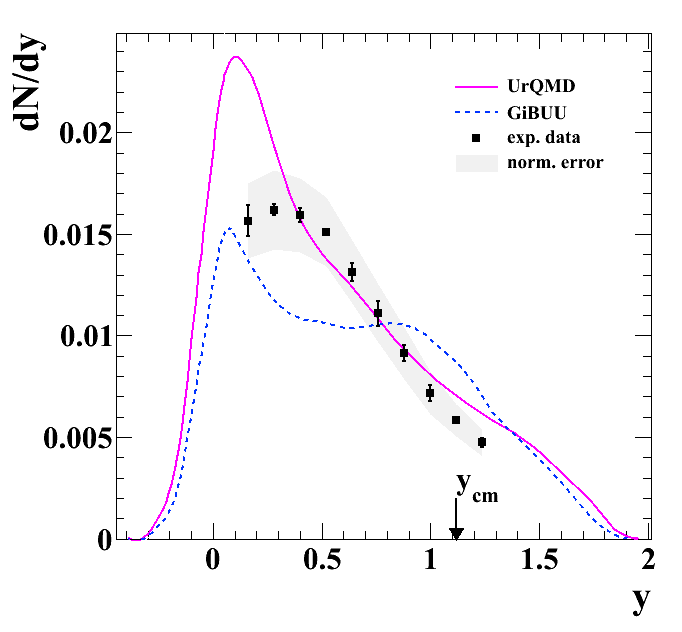}
\caption[]{Experimental rapidity-density distribution, $dN/dy$, of $\Lambda$ hyperons 
(symbols). The error bars show the systematic errors. 
The gray-shaded band represents the uncertainty due of the absolute normalization. 
The model curves and the arrow have the same meaning as in Fig.\,\protect\ref{lambda_TB}. 
\label{lambda_rapidity}}
\end{center}
\end{figure}

Integrating the two-dimensional phase-space distribution over transverse momentum, 
the $\Lambda$ rapidity-density distribution, $dN/dy$, is derived. It is displayed in 
Fig.\,\ref{lambda_rapidity}. 
Except for the first data point, the statistical errors are smaller than the symbols. 
The given error bars represent the systematic errors. 
The gray-shaded band displays the uncertainty of the absolute normalization of about 12\,\%. 
In contrast to symmetric heavy-ion collisions at comparable 
beam energies, where the rapidity distributions of secondary particles 
are symmetric around maxima at the center-of-mass rapidity of the nucleon-nucleon system 
\cite{Merschmeyer07,Justice98,hades_Lambda_ArKCl,Albergo02,Ahmad96,Barrette00}, 
the present $\Lambda$ yield at rapidities of $y>0.3$ 
decreases monotonically with increasing rapidity. 
Unfortunately, the interesting region around target rapidity is 
not covered by the detector, due to its limiting upper polar angle, $\theta<85$~degrees. 
For comparison, we note that our pion and $\eta$ meson analyses in the same 
collision system, p\,(3.5~GeV)\,+\,Nb, showed 
Gaussian shaped rapidity distribution centered, however,   
at rapidities of about 0.95 \cite{hades_pion_eta_pNb}, i.e. well below $y_{cm}$. 
Even slower emission sources ($\beta \simeq 0.5-0.6$) are estimated  
by KaoS for $K^+$ and $K^-$ production in p+Au collisions at 3.5\,GeV \cite{Scheinast06}. 
It is worth to be noted that this observation is confirmed by our preliminary 
data on $K^0$ meson production in p+Nb exhibiting a rapidity density distribution 
centered around $y\simeq 0.6$ \cite{Lapidus13}.   

Finally, we try to estimate the total production probability of $\Lambda+\Sigma^0$ hyperons. 
Since the shape of the experimental distribution at the acceptance limits is not precisely known, 
especially at large polar angles, i.e. around target rapidity, $y\sim0$, this attempt  
is possible only in a model-dependent way. Since the $dN/dy$ distribution    
is declining almost linearly, we performed a straight-line fit to the data points and 
extrapolated the distribution for $y>1.3$ linearly. Similarly, the $dN/dy$ 
distribution at low rapidities, $y<0.1$, is assumed to increase linearly from zero at $y=-0.3$ to 
the first data point. The resulting total yield amounts to $0.017\pm0.003$, 
where the error comprises both, experimental and systematic errors. With this extrapolation, 
about 20\,\% of the yield is outside of the experimentally accessible rapidity range.  
For comparison, our preliminary $K^0$ yield in p+Nb, as estimated 
from the integral of a Gaussian function fitted to the $K^0_S$ rapidity distribution 
\cite{Lapidus13}, amounts to $0.011\pm 0.002$. With the yield ratios of 
$(\Lambda + \Sigma^+ + \Sigma^0 + \Sigma^-)/(\Lambda+\Sigma^0)= 1.44$,  
$(K^+ +K^0)/K^0= 2.24$, and $K^-/K^0=0.013$ taken from UrQMD, we find 
strangeness balance, i.e. the equality of the total number of $s$ and $\bar s$ quarks, 
being nicely fulfilled on average. 

\subsubsection{Comparison with other data} \label{other_exp_data}
Looking for other experimental results on $\Lambda$ production in pA collisions, 
no data could be found at beam energies below 9\,GeV. The results next to ours 
are derived with the JINR Dubna 2\,m propane bubble chamber for collisions of 
p+C  at 10\,GeV/c \cite{Aslanyan05,Aslanyan07}, i.e. already at an excess energy w.r.t. 
the $\Lambda$ threshold in NN collisions of $\sqrt{s_{NN}}-\sqrt{s_{NN,\Lambda}}=2$\,GeV. The 
authors report a $\Lambda$ production probability of $(0.053\pm0.005)$. 
The transverse momentum spectrum was found slightly harder than ours, 
as expected from the higher beam energy allowing for more energetic particles and hence larger 
transverse momenta. The rapidity distribution appeared asymmetric with the 
upper tail reaching to rapidities of 2.6. The maximum and mean values are located at about 
0.8 and 1.0, respectively, i.e. both are well below the corresponding c.m. rapidity 
for NN collisions ($y_{cm}= 1.53$), qualitatively similar to our observation  
(Fig.\,\ref{lambda_rapidity}). Similar results have already been reported 
earlier in central collisions of 
carbon and oxygen at 4.5\,GeV/c beam momentum per incident nucleon on different target nuclei,  
as measured with the 2\,m streamer chamber SKM-200 at the Synchrophasotron in Dubna    
\cite{Anikina83,Anikina84}. Systematically increasing the target mass, 
the authors found a $\Lambda$ rapidity distribution which steadily became asymmetric 
and shifted towards target rapidity.

To be able to compare 
our $\Lambda$ production probability to the strange particle yields of another proton-nucleus 
experiment performed at the same beam energy but with different target nuclei, i.e. KaoS 
data on $K^+$ and $K^-$ production in p+C and p+Au collisions \cite{Scheinast06}, 
we normalize the yields to the number of participants, $A_{part}$. 
Using a nuclear overlap (Glauber) model \cite{Eskola89} 
(with a Woods-Saxon density profile, an inelastic nucleon-nucleon 
cross section of $\sigma_{NN}^{inel}=30$\,mb, 
and an impact parameter range of $b=0-10$\,fm), 
we get $A_{part}= 2.5~(3.3)$ for p+Nb (p+Au). Hence, we derive for our system, p+Nb,  
a normalized $\Lambda+\Sigma^0$ yield being about 1.9 times larger than the $K^+$ 
yield per number of participants for p+Au collisions at the same beam energy \cite{Scheinast06}. 

\subsubsection{Transport model predictions}\label{simulation}
Two transport approaches are compared with the experimental data, i.e. 
the Ultra-relativistic Quantum Molecular Dynamics (UrQMD) model \cite{UrQMD1,UrQMD2} and 
the Giessen Boltzmann Uehling Uhlenbeck (GiBUU) model \cite{GiBUU,Weil12}.   
For UrQMD\footnote{http://urqmd.org}, we used code release version 3.3p1, while 
for GiBUU\footnote{https://gibuu.hepforge.org}, 
we worked with release 1.6.6179 in real-particle mode (w/o mean-field baryon-baryon potential). 
In GiBUU, a threshold energy of $\sqrt{s} = (3.4\pm0.1) $\,GeV (default) 
steers, for baryon-baryon collisions, the smooth transition from the resonance model 
to the PYTHIA implementation (version 6.4.26). 
\begin{figure}[!htb]
\begin{center}
\includegraphics[width=1.\linewidth,viewport=0 0 690 990]{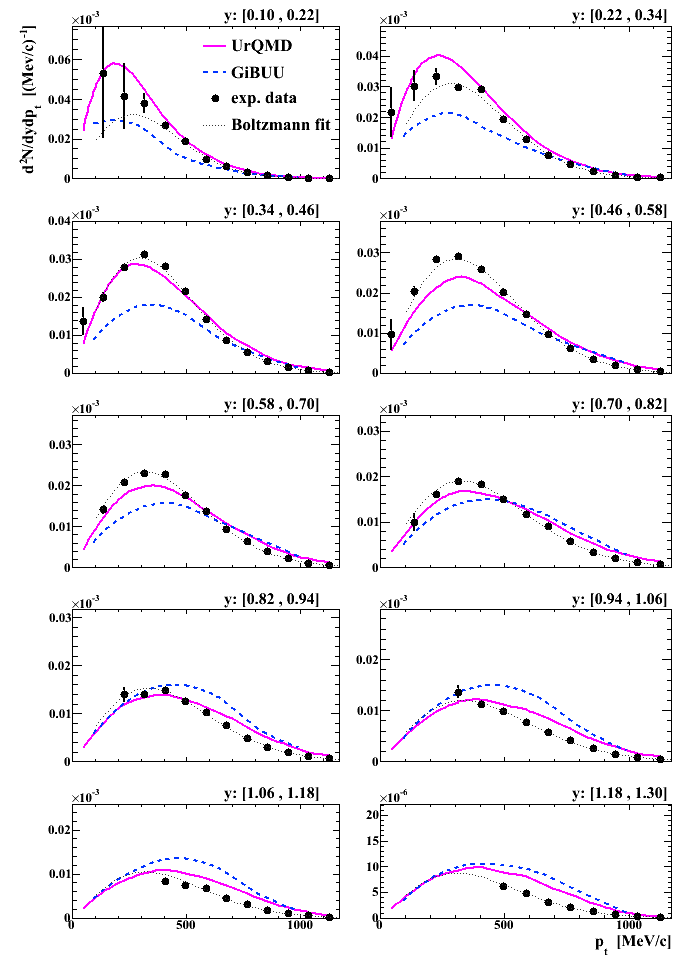}
\caption[]{Experimental $\Lambda$ transverse momentum distribution 
(symbols) for successive windows in rapidity (given on top). 
The model curves have the same meaning as in Fig.\,\protect\ref{lambda_TB}. 
The dotted curves represent Maxwell-Boltzmann distributions fitted to the data.  
\label{lambda_pt_spect_model}}
\end{center}
\end{figure}

The rapidity dependence of the inverse slope dependence presented in Fig.\,\ref{lambda_TB} 
is neither reproduced by the UrQMD nor by the GiBUU model. 
Both approaches significantly overestimate this transverse shape parameter of the 
phase-space distribution for most of the accessible rapidities. 

In contrast, the rapidity density distribution, $dN/dy$, 
presented in Fig.\,\ref{lambda_rapidity} is 
fairly well described by UrQMD, not only in shape but also on an absolute scale, over a large  
part of the experimentally accessible rapidity range, while 
GiBUU systematically underestimates (overestimates) the yield below (above) rapidities of 
$y\simeq0.8$ and does not reproduce the shape. 
The strong yield around target rapidity visible in both models, however, 
can not be compared with experimental data due to acceptance limitations. 

Inspecting the transverse-momentum distributions within rapidity slices, $d^2N/dy\,dp_t$, 
as displayed on a linear scale in Fig.\,\ref{lambda_pt_spect_model}, one realizes that the 
highest predictive power of the transport models would result from the low transverse 
momenta, where, however, the detector acceptance often prevents reliable experimental data. 
For rapidities $y\gtrsim0.3$, both models exhibit $p_t$ spectra with shapes 
similar to those of the single-slope Maxwell-Boltzmann distributions  
fitted to the experimental data (dotted curves). In the target-rapidity range, however,  
the models show deviations from the one-slope 
shape,  i.e. a superposition of differently hard spectra. The two main sources of these spectra 
are the contribution from first chance collisions yielding a rapidity spectrum centered 
around $y_{cm}$ and the contribution from subsequent hyperon-nucleon (YN) 
collisions slowing down the $\Lambda$'s down to target rapidity, $y=0$. This finding could 
be well established by inspecting the sub-processes governing the $\Lambda$ 
rapidity distributions generated with the models (cf. Fig.\,\ref{lambda_rapidity}).  
Thus, increasing in GiBUU the YN cross sections by a factor of 
two the $\Lambda$ yield could be partially redistributed to lower rapidities leading to a 
stronger enhancement around target rapidity. 
However, neither the shape nor the absolute yield of the experimental rapidity distribution 
could be reproduced. 

We tested the influence of the energy threshold at which GiBUU switches from the 
resonance model to PYTHIA in baryon-baryon collisions. 
Decreasing this threshold, hence increasing the operating range of PYTHIA, 
the inverse slope parameter $T_B(y)$ derived from Boltzmann fits to the model distributions 
decreased (i.e. the spectra got softer) 
and the rapidity density $dN/dy$ increased, for all rapidities. E.g., for a threshold of 
2.6\,GeV (default value in release 1.5), 
$T_B(y)$ from the simulation is throughout smaller 
than the experimental data with a maximum of 75\,MeV at $y\simeq0.7$, while 
the yield $dN/dy$ increases roughly by an overall factor of 1.5.   
Apparently, in PYTHIA some elementary cross sections being 
relevant for $\Lambda$ production are larger than the corresponding ones 
implemented into the resonance model. For a threshold of 3.3\,GeV, 
the average $\Lambda$ yield within the detector acceptance would match the 
experimental one. The shape of the experimental $dN/dy$ distribution, however, could not be 
reproduced, i.e. the model distribution falls slower with rapidity than the 
experimental one (cf. Fig.\,\ref{lambda_rapidity}). 

Finally, on an absolute scale, the UrQMD approach does a better job than the  
GiBUU model, since it largely reproduces the experimental 
transverse-momentum and rapidity-density distributions. The ongoing analysis  
of inclusive $K^0$ and $\Lambda$ production in pp collisions 
at the same kinetic beam energy of 3.5\,GeV will provide cross section measurements of  
quite a number of elementary reaction channels 
(especially $pp\rightarrow\Delta^{++} K^0 \Lambda/\Sigma^0$) and 
perhaps even of new channels (e.g. channels involving $\Sigma(1385)$ or $\Lambda(1405)$ 
\cite{hades_pp_Sigma1385,hades_pp_Lambda1405}) which may help to improve the 
resonance-model part of GiBUU. 

\subsection{$\mathbf{\Lambda}$ Polarization} \label{polarization}
\begin{figure}[!htb]
\begin{center}
\includegraphics[width=1.\linewidth,viewport=0 0 470 890]{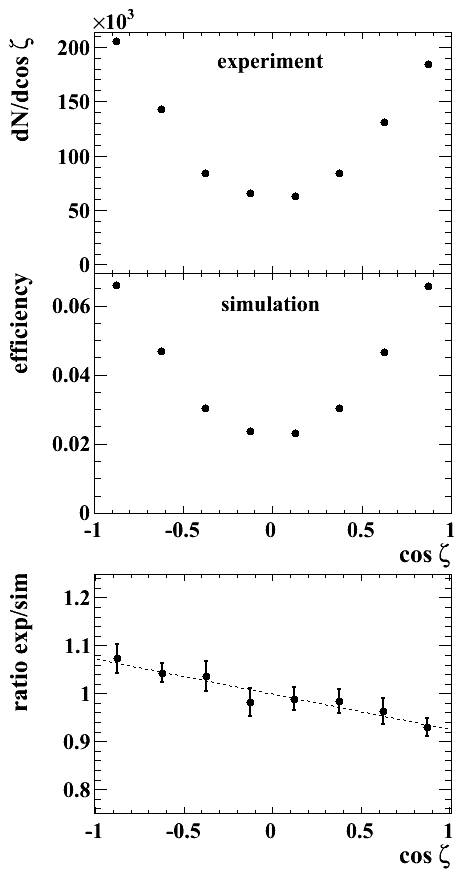}
\caption[]{Top: Raw angular distribution of the proton in the $\Lambda$ rest frame 
relative to the $\Lambda$ production plane normal. Middle: The same for polarization-free 
simulation data normalized to the corresponding input distribution. 
Bottom: Corrected and arbitrarily normalized angular distribution. The displayed error bars 
comprise both, statistical and systematic, errors. 
The dashed line is a fit according to Eq.\,(\ref{def_lambda_polarization}) 
used to extract the polarization.    
\label{lambda_pol_angle}}
\end{center}
\end{figure}
$\Lambda$ hyperons might be polarized perpendicularly to the production plane, i.e. a plane 
defined by the momentum vector of the beam and the momentum vector of the hyperon. The normal 
vector of that plane is defined as 
\begin{equation}
\vec n=\frac{\vec p_{beam}\times\vec p_{\Lambda}}{\vert\vec p_{beam}\times\vec p_{\Lambda}\vert}. 
\label{def_prod_plane}
\end{equation}
The relevant angle to be studied is the angle $\zeta$ between $\vec n$ and the 
momentum vector $\vec p_{p}^{\ast}$ of the decay proton in the $\Lambda$ rest frame 
(marked by an asterisk), i.e. 
\begin{equation}
\cos \zeta=\frac{\vec p_p^{\ast} \cdot \vec n}{\vert \vec p_p^{\ast} \cdot \vec n \vert}. 
\label{def_lambda_pol_angle}
\end{equation}
The corresponding experimental angular distribution is displayed in the upper panel of 
Fig.\,\ref{lambda_pol_angle}. 
It is derived similarly to the filling of the 
raw phase-space distribution of the $\Lambda$, i.e. performing, for each angular bin 
$d\cos \zeta$, a combinatorial-background subtraction on the proton-$\pi^-$ invariant-mass 
distribution with subsequent $\Lambda$-peak integration. 
The corresponding (polarization-free) UrQMD-simulated 
angular distribution is given in the middle panel of Fig.\,\ref{lambda_pol_angle}. 
The yield suppression around $\cos \zeta \sim 0$ is the result of the detector acceptance. 
Both, the experimental and simulated, angular distributions 
exhibit a high degree of mirror symmetry w.r.t. the $\Lambda$ production plane. 
Since the $\Lambda$ statistics is copious, 
we extract the $\Lambda$ polarization $\cal P$ by fitting a straight line 
\begin{equation}
\frac{dN}{d\cos \zeta}= C' (1+ \alpha{ \cal P} \cos\zeta) 
\label{def_lambda_polarization}
\end{equation}
to the distribution of the corrected (experimental/simulation) 
angular distribution shown in the lower panel of Fig.\,\ref{lambda_pol_angle}. 
Note that in Eq.\,(\ref{def_lambda_polarization}), the asymmetry parameter, 
$\alpha =0.642\pm0.013$, of the parity-violating weak decay of the $\Lambda$ hyperon 
is a measure of the interference between s and p waves of the final state \cite{PDG2012}. 
The $\Lambda$ polarization, averaged over the available phase space, is 
determined with different methods, thus allowing for a consistence check of the various analyses. 
Method (1) implies first the averaging over a 
set of $\Lambda$ phase-space distributions resulting from a reasonable, i.e. $\pm 20\%$, 
variation of the topological cuts to select the $\Lambda$ candidates (cf. Sect.\,\ref{lambda_pid}) 
and then one overall fit to the mean angular distribution. It delivers  
${\cal P}=-0.115 \pm 0.005 \pm 0.021$. Method (2) involves first individual angular fits 
(each for a certain geometrical cut setting) yielding a variety of polarization numbers 
and then an average (i.e. weighted with the proper yield) over this set of fit results. Here, 
we get ${\cal P}=-0.123 \pm 0.005 \pm 0.012$, in agreement with the value derived with method (1). 
In both cases, statistical and systematic errors are given, where 
the latter ones include the variations of the topological cuts and of the fit ranges 
applied to the p-$\pi^-$ invariant-mass distributions. 
Finally, method (3) makes use of the mirror symmetry mentioned above.  
Though it is possibly not fully met due to slight acceptance differences at the detector edges 
of experiment and simulation, it can be used to determine the polarization by simply 
integrating the corrected angular distribution over the positive (Up) and negative (Down) 
cosine of the angle $\zeta$, i.e. 
\begin{equation}
{\cal P} = \frac{2}{\alpha}\, \frac{\mathrm{Up-Down}}{\mathrm{Up+Down}}. 
\label{def_pol_up_down}
\end{equation}    
(Another, equivalent, approach to the polarization is based on the relation  
$\alpha {\cal P}=\langle \cos \zeta \rangle / \langle \cos^2 \zeta \rangle$, where 
the angle brackets imply the average over the entire angular range and all events.) 
The corresponding overall polarization derived with Eq.\,(\ref{def_pol_up_down}) 
amounts to $-0.104 \pm 0.008 \pm 0.027$, in agreement with the values derived with the linear fit. 
The mean value following from the various methods amounts to 
$\langle {\cal P} \rangle=-0.119 \pm0.005 \pm 0.016$   
with the given statistical and systematic errors. 
For the differential $\Lambda$ polarization investigated in the following, 
the linear fit with Eq.\,(\ref{def_lambda_polarization}) is used, and 
the systematic errors are accounted for by method (1).  

\begin{figure}[!htb]
\begin{center}
\includegraphics[width=1.\linewidth,viewport=0 0 600 600]{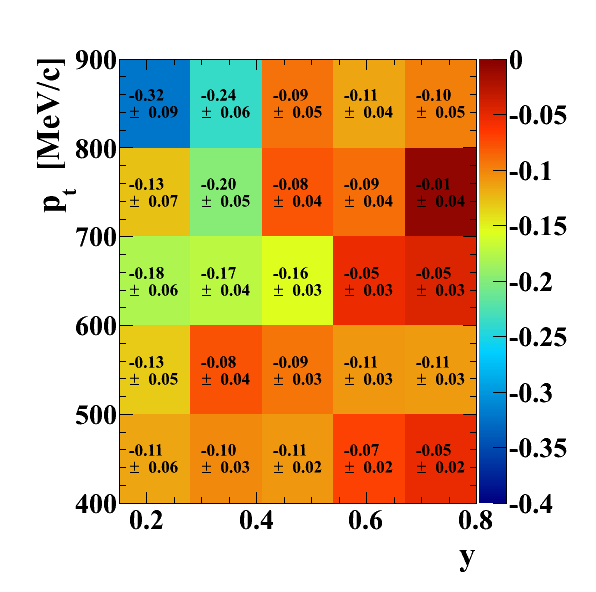}
\caption[]{$\Lambda$ polarization as a function of rapidity and transverse momentum, 
${\cal P} (y,p_t)$. The errors attached to the polarization values comprise both, 
statistical and systematic, errors. The axis on the right displays the linear color coding 
applied to the polarization values only, not involving the corresponding errors. 
\label{lambda_pol_vs_pty}}
\end{center}
\end{figure}

Now, we study the phase-space dependence of the $\Lambda$ polarization. 
Figure\,\ref{lambda_pol_vs_pty} shows the $\Lambda$ 
polarization as a function of rapidity and transverse momentum, ${\cal P}(y,p_t)$.  
Note that, for this (quasi triple-differential) figure, 
sixteen hundred invariant-mass distributions of proton-$\pi^-$ pairs 
have been analyzed (i.e., 8 topological cut sets, 
8 $\cos\zeta$ bins, 5 $p_t$ bins, 5 $y$ bins).  
We exclusively observe negative central polarizations values over the entire phase space displayed.  
Even when incorporating the given errors, this finding keeps almost unchanged. 
Though the statistics is copious, the absolute value of the polarization is somewhat fluctuating, 
i.e. it shows differences between the central values determined in neighbouring phase-space cells. 
Typically, these deviations of the polarization values from the average trends 
in slices of rapidity or transverse momentum (determined, e.g., by linear regression), 
are well within the errors. A clear tendency from the lower right to the upper left corner is found: 
The polarization is strongest at low rapidities and large transverse momenta. Its absolute value   
increases with transverse momentum for the two lowest rapidity bins while it is  
almost independent of $p_t$ within the remaining $y$ region; and it decreases with rapidity 
for the three upper transverse-momentum bins while it hardly changes with rapidity for the 
two low-$p_t$ bins.  

\begin{figure}[!htb]
\begin{center}
\includegraphics[width=1.\linewidth,viewport=0 0 630 570]{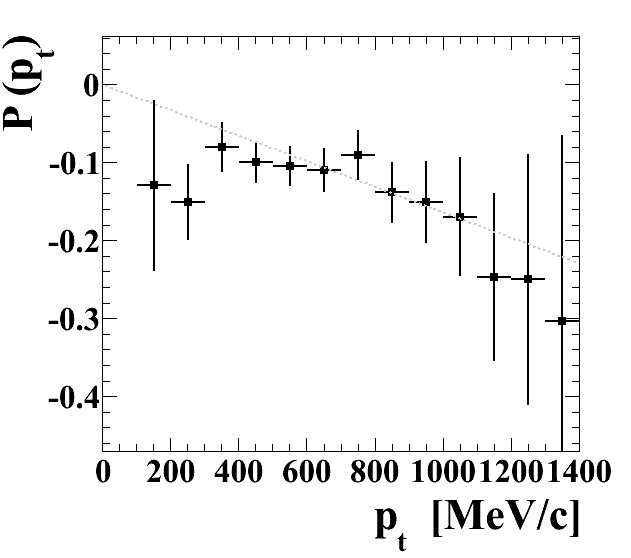}
\caption[]{$\Lambda$ polarization as a function of the $\Lambda$ 
transverse momentum, ${\cal P}(p_t)$. The vertical bars associated 
with symbols represent the combined statistical and systematic errors. 
The dotted curve is a fit to the data with a straight line (cf. text). 
\label{lambda_pol_vs_pt}}
\end{center}
\end{figure}

To make this finding more obvious, we studied the one-dimensional dependences. 
For that reason, we included also the regions outside the sharp upper and lower 
rapidity and transverse-momentum limits of Fig.\,\ref{lambda_pol_vs_pty} 
which were applied for statistical reasons. 
Integrating over all experimentally populated rapidities, 
Fig.\,\ref{lambda_pol_vs_pt} shows the dependence of the 
$\Lambda$ polarization on the $\Lambda$ transverse momentum. Its absolute value clearly increases 
with $p_t$. The dotted line is a linear fit to the data with the simplest formula ensuring a 
vanishing polarization at zero transverse momentum, i.e.  
${\cal P}(p_t) = D \, p_t$. 
The corresponding slope amounts to $D=(-0.18 \pm 0.02)~$(GeV/c)$^{-1}$. 
Looking for phase-space dependences of earlier measurements one realizes that, because of the 
fixed $\Lambda$ production angle in most of the high-energy experiments, a definite Feynman-$x$ 
value corresponds to a certain transverse momentum. Hence, there the $p_t$ dependence of the 
$\Lambda$ polarization reflects both $p_t$- and $x_F$-dependences. Nevertheless, 
our slope $D$ is well in between the $p_t$ slopes of about -0.05 to -0.30~(GeV/c)$^{-1}$ 
found for pN-reactions at energies 
from 400 to 800 GeV and presented in the compilation of ref.\,\cite{Fanti99}. 

\begin{figure}[!htb]
\begin{center}
\includegraphics[width=1.\linewidth,viewport=0 0 670 590]{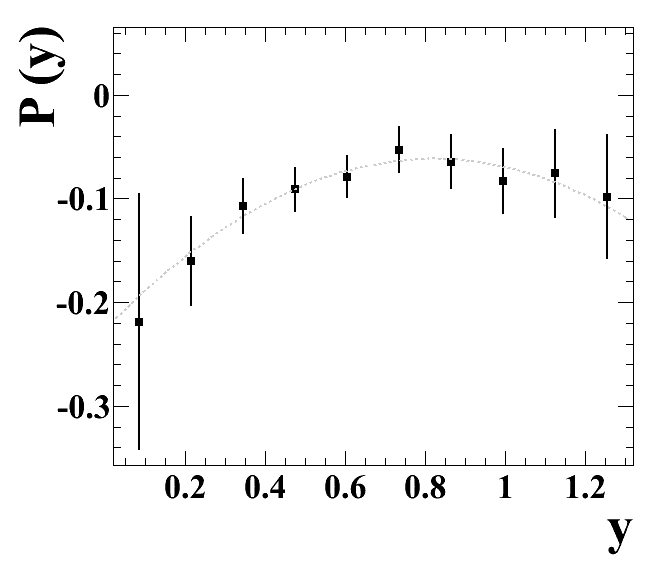}
\caption[]{$\Lambda$ polarization as a function of the $\Lambda$ rapidity$, {\cal P}(y)$. 
The vertical bars associated 
with symbols represent the combined statistical and systematic errors. 
The dotted curve is a fit to the data with a parabola (cf. text). 
The arrow indicates the center-of-mass rapidity in nucleon-nucleon collisions at 3.5\,GeV. 
\label{lambda_pol_vs_rap}}
\end{center}
\end{figure}

Integrating instead over all experimentally accessible transverse momenta, 
Fig.\,\ref{lambda_pol_vs_rap} gives the rapidity dependence of the $\Lambda$ polarization, 
${\cal P}(y)$. Its  absolute value is smallest at an intermediate rapidity, $y \simeq 0.82\pm0.09$, 
as derived from a parabolic fit (dotted curve) to the experimental data, and increases weakly 
both with decreasing and increasing rapidity, whereby the behaviour at the upper rapidities 
is compatible with a constant polarization. 
It is worth to recall that $\Lambda$ polarization in pp collisions is odd in $x_F$ 
(or ($y$-$y_{cm}$)) with negative (positive) sign in the beam (target) fragmentation region 
\cite{Felix96,Choi98,Roeder12}. In the present experiment, however, the $\Lambda$ hyperons 
carry exclusively negative polarizations while they populate mostly the target hemisphere 
(Fig.\,\ref{lambda_rapidity}). Hence, the number of $\Lambda$'s arising from primary (first chance) pN 
collisions is negligible, and the polarization we see at low rapidities is supposed to come 
from $\Lambda$'s which have scattered, which are produced by slowed-down beam protons, or which 
are produced by ``unwounded'' protons on a cluster of many nucleons from the target.   

\section{Summary}\label{summary}
Summarizing, we presented high-statistics data 
on production and polarization of $\Lambda$ hyperons produced in 
collisions of p\,(3.5~GeV)\,+\,Nb. The data were taken with HADES at SIS18/GSI. 
The $\Lambda$ phase-space distribution was compared with corresponding results of other 
experiments and with transport model predictions. The rapidity density was found to 
decrease monotonically within a rapidity window ranging from 0.3 to 1.3, i.e. 
from values above the target rapidity to values beyond the center-of-mass rapidity 
for nucleon-nucleon collisions. 
None of the employed models was able to reproduce fully the experimental distributions, i.e. 
in absolute yield and in shape. We link this observation to an insufficient 
modelling in the transport approaches of the elementary processes being relevant for 
$\Lambda$ production, rescattering and absorption. 
For the first time, the data allow 
for a genuine two-dimensional investigation of the $\Lambda$ polarization as a function 
of the large phase space covered by HADES. 
We found finite negative values of the polarization in the order of $5-20\%$   
over the entire phase space with its magnitude increasing almost linearly 
with increasing transverse momentum for $p_t>300$\,MeV/c, $-{\cal P}(p_t) = (0.18 \pm 0.02)~$(GeV/c)$^{-1}p_t$,  
and increasing with decreasing rapidity for $y < 0.8$. 
The average polarization amounts to 
$\langle {\cal P} \rangle=-0.119  \pm0.005\,\text{(stat)} \pm 0.016\,\text{(syst)}$. 
Assuming that the $\Lambda$ polarization origins from interactions of the primary beam protons   
with target nucleons, the surprise is the survival of the $\Lambda$ spin orientation in 
subsequent rescattering processes until kinetic freeze-out.  
Even if the $\Lambda$ hyperons are not produced in primary (first chance) collisions, 
they carry the information on the beam projectile. 

\begin{acknowledgement}
Useful conversations 
with W.~Eyrich, J.\,L.~Ritman and E.~Roderburg on $\Lambda$ 
polarization measurements are gratefully acknowledged. The HADES collaboration 
acknowledges the support by BMBF grants 05P09CRFTE, 05P12CRGHE, 06FY171, 06MT238~T5, 
and 06MT9156~TP5, by HGF VH-NG-330, by DFG EClust 153, by GSI TMKRUE, 
by the Hessian LOEWE initiative through HIC for FAIR (Germany), by EMMI (GSI), 
by grant GA~CR~13-067595 (Czech Rep.), by grant NN202198639 (Poland),
by grant UCY-10.3.11.12 (Cyprus), by CNRS/IN2P3 (France), by INFN (Italy),
and by EU contracts RII3-CT-2005-515876 and HP2 227431.
\end{acknowledgement}


\begin{thebibliography}{99}

\bibitem{Hartnack12}
Ch.~Hartnack, H.~Oeschler, Y.~Leifels, E.\,L.~Bratkovskaya, J.~Aichelin,
Phys. Rept. {\bf 510}, 119 (2012).

\bibitem{GiBUU} O.~Buss, T.~Gaitanos, K.~Gallmeister, H.\,van~Hees, M.~Kaskulov, 
O.~Lalakulich, A.\,B.~Larionov, T.~Leitner, J.~Weil, U.~Mosel, 
Phys. Rept. {\bf 512}, 1 (2012).

\bibitem{Fuchs04} C.~Fuchs, Progr. Part. Nucl. Phys. {\bf 53}, 113 (2004).

\bibitem{CBMphysicsBook12} B.~Friman et al., The CBM Physics Book,  
Lect. Notes Phys. {\bf 814}, 1 (2011). 

\bibitem{SengerStroebele99} P.~Senger and H.~Str\"obele, 
J. Phys. G: Nucl. Part. Phys. {\bf 25}, R59 (1999).  

\bibitem{Sturm01} C.~Sturm {\it et al.} (KaoS collaboration), 
Phys. Rev. Lett. {\bf 86}, 39 (2001).

\bibitem{Fuchs01} C.~Fuchs, A.~Faessler, E.~Zabrodin, Y.~Zheng, 
Phys. Rev. Lett. {\bf 86}, 1974 (2001).

\bibitem{Hartnack06} Ch.~Hartnack, H.~Oeschler, and J.~Aichelin, 
Phys. Rev. Lett. {\bf 96}, 012302 (2006).

\bibitem{AichelinKo85} J.~Aichelin and C.\,M.~Ko, Phys. Rev. Lett. {\bf 55}, 2661 (1985).

\bibitem{LiKo95} G.\,Q.~Li and C.\,M.~Ko, Phys. Lett. B~{\bf 349}, 405 (1995).

\bibitem{LiBrown_kpm98} G.\,Q.~Li and G.\,E.~Brown, 
Phys. Rev. C~{\bf 58}, 1698 (1998).

\bibitem{CassBrat99} W.~Cassing, E.\,L.~Bratkovskaya, Phys. Rep. {\bf 308}, 65 (1999). 

\bibitem{Kwisnia00} K.~Wisniewski {\it et al.} (FOPI collaboration), 
Eur. Phys. J. A~{\bf 9}, 515 (2000). 

\bibitem{Foerster03} A.~F\"orster {\it et al.} (KaoS collaboration),
Phys. Rev. Lett. {\bf 91},  152301 (2003).

\bibitem{Merschmeyer07} M.~Merschmeyer {\it et al.} (FOPI collaboration),
Phys. Rev. C~{\bf 76}, 024906 (2007).
 
\bibitem{Foerster07} A.~F\"orster {\it et al.} (KaoS collaboration),
Phys. Rev. C~{\bf 75}, 024906 (2007).	

\bibitem{Lotfi09} M.\,L.~Benabderrahmane {\it et al.} (FOPI collaboration), 
Phys. Rev. Lett. {\bf 102}, 182501 (2009).

\bibitem{hades_K0_ArKCl}  G.~Agakishiev {\it et al.} (HADES collaboration),
Phys. Rev. C~{\bf 82}, 044907 (2010).

\bibitem{Crochet00} P.~Crochet {\it et al.} (FOPI collaboration), 
Phys. Lett. B~{\bf 486}, 6 (2000). 

\bibitem{Uhlig05} F.~Uhlig  {\it et al.} (KaoS collaboration), 
Phys. Rev. Lett. {\bf 95}, 012301  (2005).

\bibitem{Leifels10} Y.~Leifels (FOPI collaboration), 
J. Phys. G: Conf. Ser. {\bf 230}, 012002 (2010).

\bibitem{LiKo96} G.\,Q.~Li and C.\,M.~Ko, Phys. Rev. C~{\bf 54}, 1897 (1996).

\bibitem{LiBrown98} G.\,Q.~Li, G.\,E.~Brown, Nucl. Phys. A~{\bf 636}, 487 (1998).

\bibitem{Wang99} Z.\,S.~Wang, A.~Faessler, C.~Fuchs, T.~Gross-Boelting,
Nucl. Phys. A ~{\bf 645}, 177 (1999).

\bibitem{Ritman95} J.\,L.~Ritman {\it et al.} (FOPI collaboration),
Z. Phys. A~{\bf 352}, 355 (1995).

\bibitem{Justice98} M.~Justice {\it et al.} (EOS collaboration), 
Phys. Lett. B~{\bf 440}, 12 (1998).

\bibitem{hades_Lambda_ArKCl} G.~Agakishiev {\it et al.} (HADES collaboration),
Eur. Phys. J. A~{\bf  47}, 21 (2011).

\bibitem{Scheinast06} W.~Scheinast {\it et al.} (KaoS Collaboration), 
Phys. Rev. Lett. {\bf 96}, 072301 (2006). 
	
\bibitem{Bunce76} G.~Bunce  {\it et al.}, Phys. Rev. Lett. {\bf 36}, 1113 (1976).

\bibitem{Heller77} K.~Heller, O.\,E.~Overseth, G.~Bunce, F.~Dydak, H.~Taureg, 
Phys. Lett. {\bf 68B}, 480 (1977).

\bibitem{Heller78} K.~Heller  {\it et al.}, Phys. Rev. Lett. {\bf 41}, 607 (1978).

\bibitem{Lomanno79} F.~Lomanno, D.~Jensen, M.\,N.~Kreisler, R.~Poster, J.~Humphrey,
Phys. Rev. Lett. {\bf 43}, 1905 (1979).

\bibitem{Abe83} F.~Abe {\it et al.}, Phys. Rev. Lett. {\bf 50}, 1102 (1983).

\bibitem{Smith87} A.\,M.~Smith {\it et al.}, Phys. Lett. B~{\bf 185}, 209 (1987).

\bibitem{Bonner88} B.\,E.~Bonner {\it et al.}, Phys. Rev. D~{\bf 38}, 729 (1988).

\bibitem{Fanti99} V.~Fanti {\it et al.} (NA48 collaboration), Eur. Phys. J. C~{\bf 6}, 265 (1999).

\bibitem{Lundberg89} B.~Lundberg {\it et al.}, Phys. Rev. D~{\bf 40}, 3557 (1989).

\bibitem{Felix96} J.~Felix {\it et al.} (E766 collaboration), Phys. Rev. Lett. {\bf 76}, 22 (1996).

\bibitem{Felix99} J.~Felix {\it et al.} (E766 collaboration), 
Phys. Rev. Lett. {\bf 82}, 5213 (1999).

\bibitem{Bellwied02} R.~Bellwied (for the E896 collaboration), 
Heavy Ion Physics {\bf 15}, 437 (2002).

\bibitem{Adamovich04} M.\,I.~Adamovich  {\it et al.} (WA49 collaboration), 
Eur. Phys. J. C~{\bf 32}, 221 (2004).

\bibitem{Airapetian07} A.~Airapetian  {\it et al.} (HERMES collaboration), 
Phys. Rev. D~{\bf 76}, 092008 (2007).

\bibitem{Choi98} S.~Choi (for the DISTO collaboration), Nucl. Phys. A~{\bf 639}, 1c (1998).

\bibitem{Balestra99} F.~Balestra {\it et al.} (DISTO collaboration), 
Phys. Rev. Lett. {\bf 83}, 1534 (1999).

\bibitem{Pizzolotto07}  C.~Pizzolotto, PhD thesis, 
Friedrich-Alexander-Universit\"at Erlangen-N\"{u}rnberg (2007).

\bibitem{Roeder11}  M.~R\"{o}der, PhD thesis, Ruhr-Universit\"at Bochum (2011).

\bibitem{Roeder12} M.~R\"{o}der and J.~Ritman (for the COSY-TOF collaboration), 
EPJ Web of Conferences {\bf 37}, 01008 (2012).

\bibitem{Harris81} J.\,W.~Harris, A.~Sandoval, R.~Stock, H.~Stroebele, R.\,E. ~Renfordt, 
J.\,V.~Geaga, H.\,G.~Pugh, L.\,S.~Schroeder, K.\, L.~Wolf, A.~Dacal, 
Phys. Rev. Lett. {\bf 47}, 229 (1981).

\bibitem{Felix99a} J.~Felix, Mod. Phys. Lett. {\bf 14}, 827 (1999).

\bibitem{Siebert08} H.-W.~Siebert, Eur. Phys. J. Special Topics {\bf 162}, 147 (2008).

\bibitem{Nurushev2013} S.\,B.~Nurushev, M.\,F.~Runtso, M.\,N.~Strikhanov,
Lect. Notes. Phys. {\bf 859}, 343 (2013). 

\bibitem{Andersson79} B.~Andersson, G.~Gustafson, and G.~Ingelman, 
Phys. Lett. B~{\bf 85}, 417 (1979).

\bibitem{DeGrand81} Th.\,A.~DeGrand and H.\,U.~Miettinen,
Phys. Rev. D~{\bf 24}, 2419 (1981). 

\bibitem{Zuo-tang97} L.~Zuo-tang and C.~Boros,
Phys. Rev. Lett. {\bf 79}, 3608 (1997).

\bibitem{Hui04} D.~Hui and L.~Zuo-tang, 
Phys. Rev. D~{\bf 70}, 014019 (2004).

\bibitem{Yamamoto97}  Y.~Yamamoto,K.~Kubo, and H.~Toki,
Prog. Theor. Phys. {\bf 98}, 95 (1997).

\bibitem{Kubo99}  K.~Kubo, Y.~Yamamoto, and H.~Toki,
Prog. Theor. Phys. {\bf 101}, 615 (1999).

\bibitem{Laget91} J.\,M.~Laget, Phys. Lett. B~{\bf 259}, 24 (1991).

\bibitem{Soffer92} J.~Soffer and N.\,A.~T\"ornqvist, 
Phys. Rev. Lett. {\bf 68}, 907 (1992).

\bibitem{Sibirtsev98} A.~Sibirtsev and W.~Cassing, e-Print: nucl-th/9802019 (1998).

\bibitem{Gasparian01} A.\,M.~Gasparian, J.~Haidenbauer, C.~Hanhart, L.~Kondratyuk, J.~Speth,
Nucl. Phys. A~{\bf 684}, 397 (2001).

\bibitem{Sibirtsev06} A.~Sibirtsev, J.~Haidenbauer, H.-W.~Hammer, and S. Krewald,
Eur. Phys. J. A~{\bf 27}, 269 (2006).

\bibitem{hades_spectro} G.~Agakichiev {\it et al.} (HADES collaboration),
Eur. Phys. J. A~{\bf 41}, 243 (2009).

\bibitem{HADES-PRL07} G.~Agakichiev {\it et al.} (HADES collaboration),
Phys. Rev. Lett. {\bf 98}, 052302 (2007).

\bibitem{PhD_Schmah} A.~Schmah, PhD thesis, Techn. Universit\"at Darmstadt (2008).

\bibitem{hades_kpm_phi} G.~Agakishiev {\it et al.} (HADES collaboration),
Phys. Rev. C~{\bf 80}, 025209 (2009).

\bibitem{hades_Xi} G.~Agakishiev {\it et al.} (HADES collaboration),
Phys. Rev. Lett. {\bf 103}, 132301 (2009).

\bibitem{hades_epm_pNb} G.~Agakishiev {\it et al.} (HADES collaboration),
Phys. Lett. B~{\bf 715}, 304 (2012).

\bibitem{hades_pion_eta_pNb} G.~Agakishiev {\it et al.} (HADES collaboration),
Phys. Rev. C~{\bf 88}, 024904 (2013).

\bibitem{Tlusty_Bormio_2012} P.~Tlusty  {\it et al.} (HADES collaboration), 
Proc. 50th Int. Winter Meeting on Nucl. Phys., Bormio (Italy) 2012, Proceedings of Science, 
PoS(Bormio2012)019.


\bibitem{PDG2012} J.~Beringer {\it et al.}, (Particle Data Group),
Phys. Rev. D~{\bf 86}, 010001 (2012).

\bibitem{Dukes87} E.~Dukes {\it et al.}, Phys. Lett. B~{\bf 193}, 135 (1987). 

\bibitem{UrQMD1} S.\,A.~Bass {\it et al.}, Prog. Part. Nucl. Phys. {\bf 41}, 255 (1998).

\bibitem{UrQMD2} M.~Bleicher {\it et al.}, J. Phys. G~{\bf 25}, 1859 (1999).

\bibitem{GEANT} GEANT 3.21, http://consult.cern.ch/writeup/geant/ (1993).
 	
\bibitem{Pluto}
I.~Fr\"ohlich {\it et al.}, Proceedings of Science, PoS(ACAT)076 (2007).

\bibitem{Wendisch14} C.~Wendisch, PhD thesis, Techn. Universit\"at  Dresden (2014).

\bibitem{Arnold13} O.~Arnold, Diploma thesis, Techn. Universit\"at  M\"unchen (2013).

\bibitem{Albergo02} S.~Albergo {\it et al.} (E896 collaboration), 
Phys. Rev. Lett. {\bf 88}, 062301 (2002).

\bibitem{Ahmad96} S.~Ahmad  {\it et al.} (E891 collaboration), 
Phys. Lett. B~{\bf 382}, 35 (1996).

\bibitem{Barrette00} J.~Barrette {\it et al.} (E877 collaboration), 
Phys. Rev. C~{\bf 63}, 014902 (2000).

\bibitem{Lapidus13}
K.~Lapidus, Proc. ``FAIRNESS 2013'', 16-21 Sep 2013, Berlin, Germany, 
J. Phys.: Conf. Ser., in print. 

\bibitem{Aslanyan05} P.\,Zh.~Aslanyan, AIP Conf. Proc. {\bf 796}, 184 (2005).

\bibitem{Aslanyan07} P.\,Zh.~Aslanyan, V.\,N.~Emelyanenko, G.\,G.~Rikhkvitzkaya, 
Phys. Part. Nucl. Lett. {\bf  4}, 60 (2007).

\bibitem{Anikina83} M.~Anikina {\it et al.}, 
Phys. Rev. Lett. {\bf 50}, 1971 (1983).

\bibitem{Anikina84}  M.~Anikina {\it et al.}, 
Z. Phys. C ~{\bf 25}, 1 (1984).

\bibitem{Eskola89} K.\,J.~Eskola, K.~Kajantie, J.~Lindfors, 
Nucl. Phys. B~{\bf 323}, 37 (1989).

\bibitem{Weil12} J.~Weil, H. van Hees, and U.~Mosel,  
Eur. Phys. J. A {\bf 48}, 111 (2012).



\bibitem{hades_pp_Sigma1385}  G.~Agakishiev {\it et al.} (HADES collaboration),
Phys. Rev. C~{\bf 85}, 035203 (2012).

\bibitem{hades_pp_Lambda1405}  G.~Agakishiev {\it et al.} (HADES collaboration),
Phys. Rev. C~{\bf 87}, 025201 (2013).

\end{thebibliography}
\end{document}